# Efficient Time-Resolved Pressure Estimation by Sparse Sensor Optimization and Non-Time-Resolved PIV

**Neetu Tiwari[1], Ajit Kumar Dubey[2]**

[1]Department of Mechanical and Aerospace Engineering, Indian Institute of Technology, Hyderabad, India

[2]Department of Mechanical and Industrial Engineering, Indian Institute of Technology, Roorkee, India

*Corresponding author: N. Tiwari, E-mail: neetu.tiwari@mae.iith.ac.in

**Abstract**

Pressure field estimation from PIV data has been a well-established technique. However, time-resolved pressure estimation strongly depends on the temporal resolution of the PIV measurements. Generally, PIV data has limited time resolution creating challenges to understand high Reynolds number flows. To overcome this challenge, sensor data measured at few optimized locations with higher time resolution is combined with PIV data using data driven methods to reconstruct time resolved velocity fields. In this context, if we wish to estimate time resolved pressure fields from non-time resolved PIV data, there are two possible approaches. Approach 1: reconstruct time-resolved velocity field first from non-time resolved PIV data using sensor data, and then time-resolved pressure fields are estimated from time-resolved pressure fields by applying pressure Poisson equation. Approach 2: first estimate non-time resolved pressure fields from non-time resolved velocity field measurements using pressure Poisson equation and then reconstruct time resolved pressure fields directly from non-time resolved pressure fields. These two approaches are compared in this study. These approaches are demonstrated for actual PIV data of flow over a cylinder. Time-resolved PIV measurements are down-sampled to mimic non-time-resolved velocity data. It was found that the approach two is approximately thirty times faster than approach one when time resolution is improved from 1 Hz to 50 Hz. This is expected because in the second approach the pressure Poisson equation needs to be solved only with non-time resolved velocity fields which reduces the computational load.

# 1. Introduction

Pressure field estimation from PIV data is a well-known technique. The pressure field can be estimated using PIV data in the Navier-Stokes or the pressure Poisson equation. In experiments, the temporal resolution of the estimated pressure field largely depends on the time resolution of the PIV data, which is determined by the sampling rate of high-speed cameras. PIV measurement with sufficient spatial resolution often lacks temporal resolution compared to point measurement techniques such as hot wire anemometry (HWA), laser Doppler anemometry (LDA) or piezoelectric pressure sensors. Due to this, a PIV measurement is often applied to studies on the spatial structure of the flow. On the other hand, the evolution of flow in time is mostly studied with point measurements due to its good time resolution capability.

For the time resolution improvement of PIV data, most of the published past study are based on POD and the time resolved PIV data is reconstructed corresponding to dominant POD modes. Temporal resolution improvement in the PIV measurement from non-time-resolved data was suggested by Legrand et al.[1]. They reconstructed phase averaged field with a correlation-based method based on POD. Integration of non-time-resolved data of PIV with time-resolved pressure data by POD and linear stochastic estimation for the temporal resolution improvement was performed by Murray et al.[2]. Linear stochastic estimation captures the correlation between flow and measurement signal. The accuracy of using single- and multi-time-delay in LSE-POD methods was investigated by Durgesh and Naughton 2010 [3]. They integrated time-resolved pressure sensor data with non-time-resolved PIV data. Later, the sparse time-resolved sensor outputs are de-noised by the Kalman filter and the Kalman smoother [4]. Their work also included virtual sensor data in time resolution improvement of PIV by linear stochastic estimation method. This method requires linearity or weak nonlinearity of the dynamics constructed by outputs of the sensors selected manually and a strong relationship between the sensor output and flow fields POD modes. Chen and co-workers [5] proposed a filtering based on three-sigma rule in extended POD method for time-resolved velocity and pressure field estimation from a non-time-resolved PIV data. Extended POD modes were evaluated by projection of the PIV snapshot matrix on the temporal basis of sensor data. When all POD modes of probe data are considered, this technique becomes equivalent to linear stochastic estimation [6,7]. Even though these studies reconstructed time resolved PIV data based on dominant POD modes, in selection of sensors location in the flow field, POD modes do not play any role. For line sensors which can measure velocity along a line like UVP (ultrasonic velocity profiler), an optimization technique for line selection has been also proposed by extending point sensor selection methods [8]. Recently, sensor selection method has been used with extended POD method for time resolution improvement of flow fields with line sensors [9]. This technique was demonstrated in PIV and ultrasonic velocity profiler's data. Later sensor optimization with QR method is demonstrated only in PIV data [10]. However, effect of sensor optimization in pressure field estimation is not yet investigated.

A sensor selection method provides the optimized sensor locations from flow fields to aid in flow control. The original interest to develop sensor selection method arose from the need to control the high dimensional systems using a minimal number of sensors in optimal locations. Optimization of sensors location for effective reconstruction by a convex relaxation method [11], a QR-pivoting-based sparse sensor placement method [12–14], and a determinant-based greedy sensor selection method [15] were proposed. These techniques use dominant POD modes and provide sensor locations for optimal reconstruction of truncated flow. The discrete empirical interpolation method (DEIM) provides [16] interpolation points for known POD features for the reconstruction of nonlinear-function distributions in reduced-order models. Manohar et al. 2018 [13] had extended the QR-based DEIM method (SSPOR) and succeeded in reducing reconstruction error when the number of sensors exceeded the number of POD modes [17]. The D-optimality-based greedy (DG) algorithm is mathematically equivalent to the SSPOR method when the number of sensors is less than that of POD modes. However, the DG algorithm is more accurate than the SSPOR method in the case with a larger number of sensors compared with POD modes [15]. This method was utilized to introduced sparse processing PIV (SPPIV) which uses sparse sensor data to obtain velocity distribution by applying sensor selection method (Kanda et al. 2021).

In the present study, time resolved pressure fields are estimated from non-time resolved PIV data using two approaches and a comparison of approaches is discussed. In the first approach, time resolution of velocity fields from PIV was improved using the DG-LSE methods and the time-resolved velocity field thus obtained was utilized for pressure field estimation using pressure Poisson equation. In the second approach, the non-time-resolved pressure field is directly evaluated from the non-time-resolved PIV data. The sensor-selection method was then applied to evaluate the optimized locations of the sensors from non-time-resolved pressure fields thereby eliminating the need of actual sensors during location optimization. Then, time-resolved point sensors data was collected only from the optimized locations. The time-resolved pressure data at the point and non-time-resolved pressure field data were combined for time-resolved pressure field estimation. To our knowledge, this integration of non-time-resolved spatial pressure data with sparse, optimized time-resolved sensors for the reconstruction of time-resolved pressure fields has not been demonstrated previously. The DG-LSE method was utilized for both the sensor location evaluation and reconstruction of the time-resolved pressure fields. Lower number of pressure Poisson equation solutions are needed in the second approach, making it faster but the relative accuracy of two approaches depends on number of sensors. In section 2 experimental setup is discussed which was used to get the PIV data of flow behind cylinder. In section 3 theoretical background of sensor selection method, pressure field estimation and both approaches are detailed. In section 4 results are presented and discussed followed by conclusions in section 5.

## 2. Experimental setup

A schematic of the experimental arrangement with relevant dimensions is shown in Fig. 1. The velocity measurement in the flow behind the cylinder (diameter of 40 mm) is conducted in the towing tank using PIV. The upstream velocity is 30 mm/s. For the PIV measurement, a high-speed camera with a 300 mm Nikkor lens and a green laser were used. A laser beam was expanded to a light sheet, and the flow was illuminated up to three times diameter downstream of the cylinder. The camera was fixed, and the images of the flow behind the cylinder in the *x-y* plane were recorded. A camera and laser system were towed together with the cylinder for the measurement. The towing carriage system was driven by a linear slider, and its speed and acceleration were controlled by the computer. Further details of the setup and measurements performed can be found in our earlier work [20–22].

For the PIV measurement, spherical tracer particles were mixed in the water. These particles are made of high-porous polymer with hydrophilic surface property to aqueous solution. The density and the diameter of the particles are 990 kg/m$^3$ and 50-120 μm, respectively. These particles are manufactured by HPSS20, Mitsubishi Chemical, Ltd. A green laser (Roithner Lasertechnik GmbH, Vienna, Austria) of the wavelength of 532 nm and the power of 30 mW was used for illumination. The flow images were recorded by a high-speed camera, FastCam Mini AX-50, Photron, made in Germany. The nikkor lens of 105 mm, f-number 2.8 was used. The size of the images was 1024×1024 pixels. Images were captured at 50 Hz, which is sufficient for time-resolved measurements at the concerned Re of 900. The vortex shedding frequency is 1.5 Hz, and for time resolve measurement, as per the Nyquist frequency criteria, the sampling frequency should be more than 3.0 Hz, which corresponds to the time interval of 0.33 sec between two snapshots. In the present case, the sampling frequency is 50Hz, and the corresponding time interval is 0.02 sec. The non-time-resolved data were obtained by downsampling the measured data to 1Hz to mimic non-time-resolved PIV measurement.

Post-processing of images was performed by PIVLAB software [23]. An FFT-based cross-correlation algorithm was used with step-wise rectangular interrogation window areas [24] from 128×64, 64×32, and 32×16. The spatial resolutions of velocity in both the *x* and *y* directions were 0.2 mm.

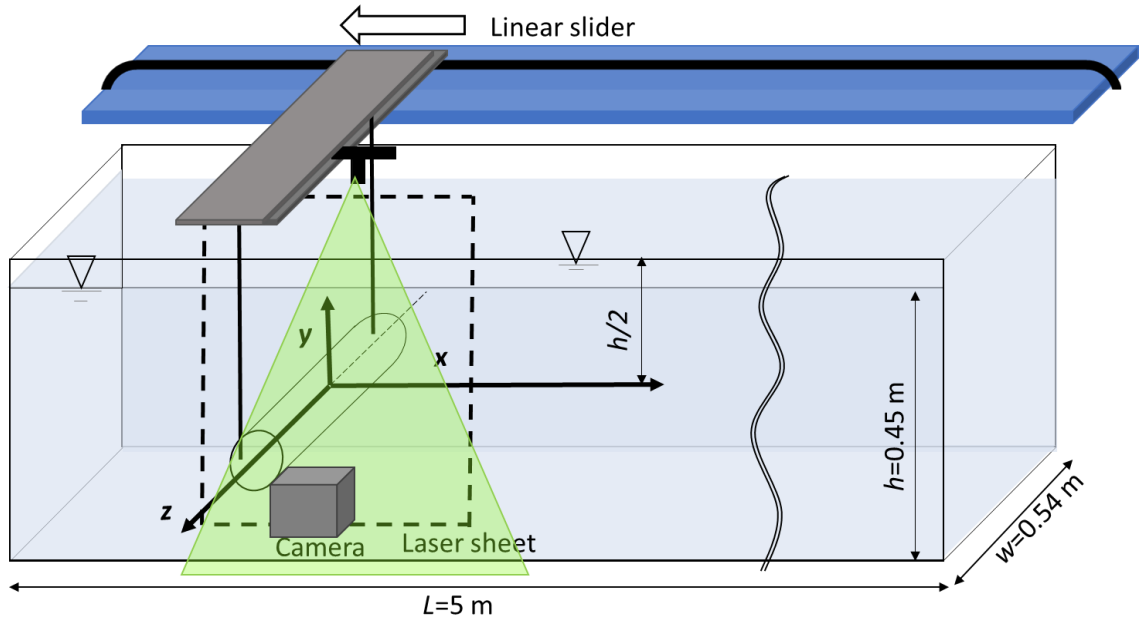

Fig. 1 Experimental arrangement for velocity measurement by PIV in flow behind a circular cylinder [11]

The methodology proposed in this study is demonstrated using experimental time-resolved PIV data of flow over a circular cylinder. From this dataset, we generated a non-time-resolved velocity field by systematic downsampling. In parallel, we extracted time-resolved pressure values at specific spatial locations to mimic the availability of sparse point pressure measurements, similar to those obtained from physical pressure sensors. These sensor signals were used in combination with the non-time-resolved PIV data to estimate time-resolved pressure fields using the proposed sensor selection based framework. The reconstructed pressure fields were then quantitatively compared with the reference pressure fields obtained directly from the full time-resolved PIV velocity data (by pressure Poisson equation). This comparison provides a direct validation of the proposed approach using experimental data.

## 3. Theoretical background
### 3.1 Sensor selection methods

Sensor selection methods based on the POD usually work on the concept that if a dynamic system can be effectively represented by the limited number of dominant POD modes, a full-state system can be reconstructed by a limited number of optimally placed sensors. Generally, time-resolved PIV data are not available for all types of flows, so the evaluation of sensor locations should be done only by using non-time-resolved data in sensor selection method.

The basic idea of the sensor selection model (Fig.2) can be written as below:

$$y_{ntr} = H\, U_{ntr}\, z_{ntr} \\ = C\, z_{ntr} \qquad (1)$$

Here $y_{ntr}$, $H$, $U_{ntr}$, and $z_{ntr}$ are the observation vector consisting of sensor values, the sparse sensor location matrix, bases of the spatial POD modes matrix, latent state variable or parameter vector corresponding to the POD coefficients respectively. $C = H\, U_{ntr}$, is the sensor candidate matrix which is to be found. Every pixel of a flow field's image is a possible sensor location, and among them, only few locations should be selected. The subscript "$tr$" and "$ntr$" stand for time-resolved and non-time-resolved data, respectively. The $H$ matrix consists of ones at the sensor location index and zeroes otherwise.

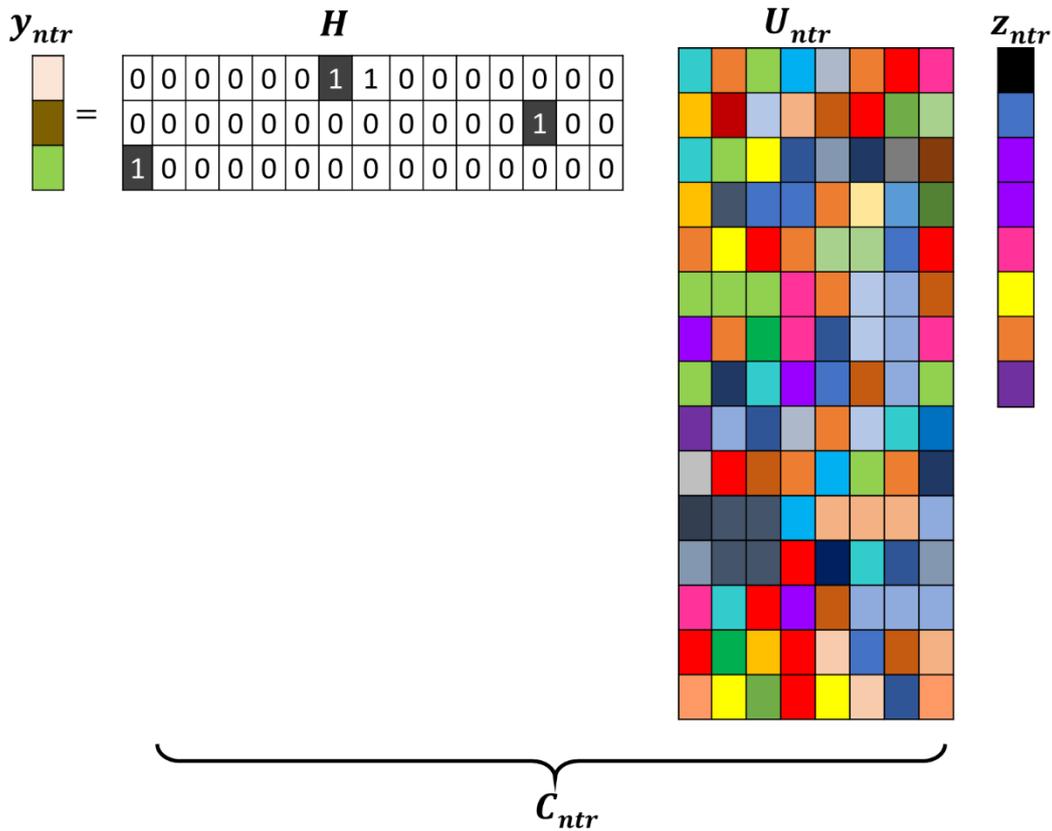

Fig. 2 Graphical representation of Eq. 1

Spatial POD modes used as bases ($U_{ntr}$) can be determined by performing the singular value decomposition (SVD) of the PIV dataset. The PIV measurement at each instant is rearranged as a single-column vector. The data $X_{ntr}$ comprises of non-time- resolved PIV measurement over time, and its SVD can be written as below:

$$X_{ntr} = U_{ntr} \Sigma_{ntr} Y_{ntr}^T \quad (2)$$

Here $\Sigma_{ntr}$ is the diagonal matrix whose entries correspond to singular values, and $Y$ is the matrix which has temporal modes. The data matrix can be written as sum of considered (*1:r*) and neglected (*r+1:N*) POD modes as below:

$$X_{ntr} = U_{ntr(1:r)}\Sigma_{ntr(1:r)}Y_{ntr(1:r)}^T + U_{ntr(r+1:N)}\Sigma_{ntr(r+1:N)}Y_{ntr(r+1:N)}^T \quad (3)$$

Here, $N$ is the total number of snapshots. The number of POD modes considered to represent the data is $r$. $X_{ntr(1:r)}$ can be evaluated by considering the first r columns of $U_{ntr}$ and $Y_{ntr}$ as follows:

$$X_{ntr(1:r)} = U_{ntr(1:r)}\Sigma_{ntr(1:r)}Y_{ntr(1:r)}^T \quad (4)$$

The truncated $X_{ntr(1:r)}$ ignores the second term, $U_{ntr(r+1:N)}\Sigma_{ntr(r+1:n)}Y_{ntr(r+1:N)}^T$ of Eq. (3).

Next, we consider an approximation of state variable, where data is represented using only a few POD modes and measurements at optimized sensor locations as below:

$$\hat{z}_{ntr} = C_{ntr}^{\dagger} y_{ntr} \quad (5)$$

With

$$C_{ntr} = H\, U_{ntr(1:r)} \quad (6)$$

The subscript † denotes the Moore-Penrose pseudoinverse of the matrix, which is calculated by "*pinv*" function of MATLAB. $\hat{z}_{ntr}$ represents approximation of $z_{ntr}$. The aim now is to evaluate $H$ and consequently $C_{ntr}$.

In the typical sensor selection problem, to check the accuracy of the placements of sensors, the sensor data is then utilized to reconstruct the same data from which the sensor is selected. In present study to reconstruct the time-resolved flow field the time-resolved sensor data were utilized. Determinant Greedy- least square estimation (DG-LSE) method is used to evaluate the sensor locations. Reconstruction of flow fields in DG method is based on a linear least squares method.

## 3.2 Determinant Greedy- least square estimation (DG-LSE) sensor selection method for time resolution improvement

In DG-LSE method, the determinant of the sensor candidate matrix is maximized to find the optimal sensor positions. The optimization problem can be expressed as below for two cases when the number of sensors ($p$) is less than or equal to POD modes ($r$) and when $p$ is greater than $r$.

$$\text{maximize } f_{DG}$$
$$f_{DG} = \begin{cases} \det(C_{ntr} C_{ntr}^T), & p \leq r \\ \det(C_{ntr}^T C_{ntr}), & p > r \end{cases} \tag{7}$$

The maximization of $f_{DG}$ is equivalent to minimization of determinant of covariance matrix of error [15]. $C_{ntr}$ as defined in Eq. 1 is the sensor candidate matrix for $p$ sensors. $C_{ntr(j)}$ is subset of $C_{ntr}$ in a sense that as position of each subsequent sensor is evaluated the size of $C_{ntr(j)}$ increases and as $p$ sensors are evaluated it becomes $C_{ntr}$. So, $C_{ntr(j)}$ represents the sensor candidate matrix during evaluation of $j^{th}$ sensor. The detailed algorithm and related derivations of DG-LSE can be referred from original work [12].

$$C_{ntr(j)} = \left[ u_{i_1}^T \; u_{i_2}^T \; \ldots \ldots \; u_{i_{j-1}}^T \; u_{i_j}^T \right]^T \tag{8}$$

Here $i_j$ is the index of $j^{th}$ selected sensor, $u_i$ is row vector of sensor candidate matrix and $u_{i_j}^T$ is the transpose of corresponding row vector of sensor candidate matrix for evaluation of $j^{th}$ sensor. The expanded form of determinant of $C_{ntr(j)} C_{ntr(j)}^T$ for $j^{th}$ sensor evaluation can be written as below:

$$\det(C_{ntr(j)} C_{ntr(j)}^T) = \det\left( \begin{bmatrix} C_{ntr(j-1)} \\ u_i \end{bmatrix} \begin{bmatrix} C_{ntr(j-1)}^T & u_i^T \end{bmatrix} \right) \tag{9}$$
$$= u_i \left( I - C_{ntr(j-1)}^T \left( C_{ntr(j-1)} C_{ntr(j-1)}^T \right)^{-1} C_{ntr(j-1)} \right) u_i^T \times \det(C_{ntr(j-1)} C_{ntr(j-1)}^T)$$

The detailed derivation of expanded form can be found in [15,25]. The greedy sensor selection method assumes that the first $(j-1)^{th}$ sensors are already known then $j^{th}$ sensor location can be evaluated by maximizing the determinant of sensor candidate matrix which leads to the following expression:

$$i_{j\_DG} = \underset{i \in \varphi \setminus \varphi_j}{\arg\max} \; u_i \left( I - C_{ntr(j-1)}^T \left( C_{ntr(j-1)} C_{ntr(j-1)}^T \right)^{-1} C_{ntr(j-1)} \right) u_i^T \tag{10}$$

Here $\varphi$ is the set of indices of sensor candidates and $\varphi_j$ is subset of indices of determined sensors.

Similarly for the case when $p > r$

$$\det(C_{ntr(j)}^T C_{ntr(j)}) = \det\left(\begin{bmatrix} C_{ntr(j-1)}^T & u_i^T \end{bmatrix} \begin{bmatrix} C_{ntr(j-1)} \\ u_i \end{bmatrix}\right) \quad (11)$$

$$\det(C_{ntr(j)}^T C_{ntr(j)}) = \left(1 + u_i(C_{ntr(j-1)}^T C_{ntr(j-1)})^{-1} u_i^T\right) \det(C_{ntr(j-1)}^T C_{ntr(j-1)})^{-1}$$

$$i_{j\_DG} = \underset{i \in \varphi \setminus \varphi_j}{\arg\max} \det\left(1 + u_i(C_{ntr(j-1)}^T C_{ntr(j-1)})^{-1} u_i^T\right) \quad (12)$$

The complete algorithm to determine the optimized location of sensors can be find in the literature [15]. In the present study, for evaluation of sensors optimized locations DG method suggested by[15] was used as it is and for time resolution improvement of PIV from time resolved sensors DG-LSE is slightly modified as discussed in section in 3.1.2. Once locations of sensors are known, the time-resolved sensor measurement should be taken at those optimized locations. In the present study, to mimic time-resolved sensor data at points, the data at optimized locations are extracted from the time-resolved flow fields (PIV measurement) by projecting the sensor location matrix onto the time-resolved data as below:

$$y_{tr\_DG} = H\, X_{tr} \quad (13)$$

The time-resolved POD coefficient can be evaluated by the following:

$$\widehat{z_{tr\_DG}} = C_{ntr}^\dagger y_{tr\_DG} = \begin{cases} C_{ntr}(C_{ntr} C_{ntr}^T)^{-1} y_{tr\_DG}, & p \leq r \\ (C_{ntr}^T C_{ntr})^{-1} C_{ntr}^T y_{tr\_DG}, & p > r \end{cases} \quad (14)$$

The time resolved flow field can now be estimated by projecting ($\widehat{z_{tr\_DG}}$) on $U_{ntr(1:r)}$.

$$\widehat{X_{tr\_DG}} = \begin{cases} U_{ntr(1:r)} C_{ntr} (C_{ntr} C_{ntr}^T)^{-1} y_{tr\_DG}, & p \leq r \\ U_{ntr(1:r)} (C_{ntr}^T C_{ntr})^{-1} C_{ntr}^T y_{tr\_DG}, & p > r \end{cases} \quad (15)$$

Since time-resolved flow fields are available, we can compare the estimated flow field with the original time-resolved flow field to assess the capability of the DG-LSE.

### 3.3 Approach 1: estimation of time-resolved velocity (by DG-LSE method) and pressure fields

The published POD based PIV time resolution improvement algorithms [4,7,26] use the following strategy:

$$X_{ntr(1:r)} = U_{ntr(1:r)} z_{ntr} \tag{16}$$

$$\hat{X}_{tr} = U_{ntr(1:r)} z_{tr} \tag{17}$$

Here $z_{ntr}$ is POD coefficient corresponding to $r$ POD modes. The non-time-resolved PIV data can be decomposed using POD in to POD modes and POD coefficients. If we can evaluate time-resolved POD coefficient ($z_{tr}$) by any means we can reconstruct $\hat{X}_{tr}$ (time-resolved PIV data) by using Eq. 17. In linear stochastic estimation-based algorithms [3,4], the correlation matrix $A$ in Eq. (18), provides the correlation between non-time-resolved PIV data with non-time-resolved POD coefficient.

$$z_{ntr} = A\, y_{ntr} \tag{18}$$

Here $y_{ntr}$ is downsampled sensor data which were measured at same instance as PIV. This $A$ Matrix can be evaluated in several ways. For example in DG sensor selection method $A = y_{ntr}^{\dagger} z_{ntr}$. Once matrix $A$ is known time resolved POD coefficient and $\hat{X}_{tr}$ can be evaluated by Eq. 19 and 17 respectively.

$$\widehat{z_{tr}} = A\, y_{tr} \tag{19}$$

Here $y_{tr}$ is time resolved sensor data. Different algorithms use different methodology to estimate $\widehat{z_{tr}}$ [5,7]. However, past studies have not addressed the importance of sensors locations on time resolution improvement of PIV data in their study except author's recent work[9]. In time resolution improvement of PIV, inclusion of sensor selection method which provides optimized locations of sensors is important and the main advantage of this inclusion is that sensors are specially tailored for dominant flow structures for which time resolved reconstruction is required.

In typical sensor selection methods, the optimized locations of sensors were obtained by using the dataset of flow field. Accuracy of optimized sensors locations was evaluated by reconstructing the original flow fields from those optimized sensors data. In other words, in the typical sensor selection methods, the data to evaluate optimized sensor locations and the data to be reconstructed were the same. The purpose of a typical sensor selection method is the evaluation of the best possible sensors positions from flow fields which could be used for flow control applications later. The typical DG sensor selection problem can be characterized in two parts as shown in Fig. 3(a); one in which sensor locations are evaluated using the DG sensor selection method and in the second part, flow field reconstruction to check the accuracy of placed sensors. So, there is no scope to increase the time resolution of non-time-resolved data. On the other hand, in the present study, for the time resolution improvement of PIV data, the sensor location evaluation and flow reconstruction parts are combined with time resolved sensors data in DG-LSE algorithm as discussed below.

Optimized sensor locations evaluation was done using only non-time resolved PIV data. After obtaining those optimized sensors locations the physical sensors are placed at those locations and time resolved point measurement are obtained. For the reconstruction of time-resolved flow fields, time resolved sensors data (obtained by measurement) are used to evaluated the time resolved POD coefficients from Eq. 14. POD modes evaluated from non-time-resolved data is further used with time resolved POD coefficients to reconstruct the flow fields using Eq. 15. Figure 3 (b) shows the present algorithm for time resolution improvement of PIV data which uses non-time-resolved velocity data of PIV with the sensor data.

The steps of the DG-LSE method are described in steps below:

1. Arrange the non-time-resolved PIV data matrix, perform singular value decomposition, and evaluate $U_{ntr}$ (Eq.3).
2. Evaluate optimized sensors locations from Eq. 10 or Eq. 12 (depending on number of sensors) using DG-LSE.
3. Measure the non-time-resolved PIV snapshots and time-resolved data at optimized sensors locations simultaneously.
4. Evaluate time-resolved POD coefficients from Eq. 14, using DG-LSE.
5. Reconstruct time-resolved velocity field data by projecting $U_{ntr}$ to $\widehat{z_{tr}}$ as given in Eq. 15, using DG-LSE.
6. Reconstruct time-resolved pressure field using time-resolved velocity field by applying the pressure Poisson equation/ Navier Stokes equation (section 3.5).

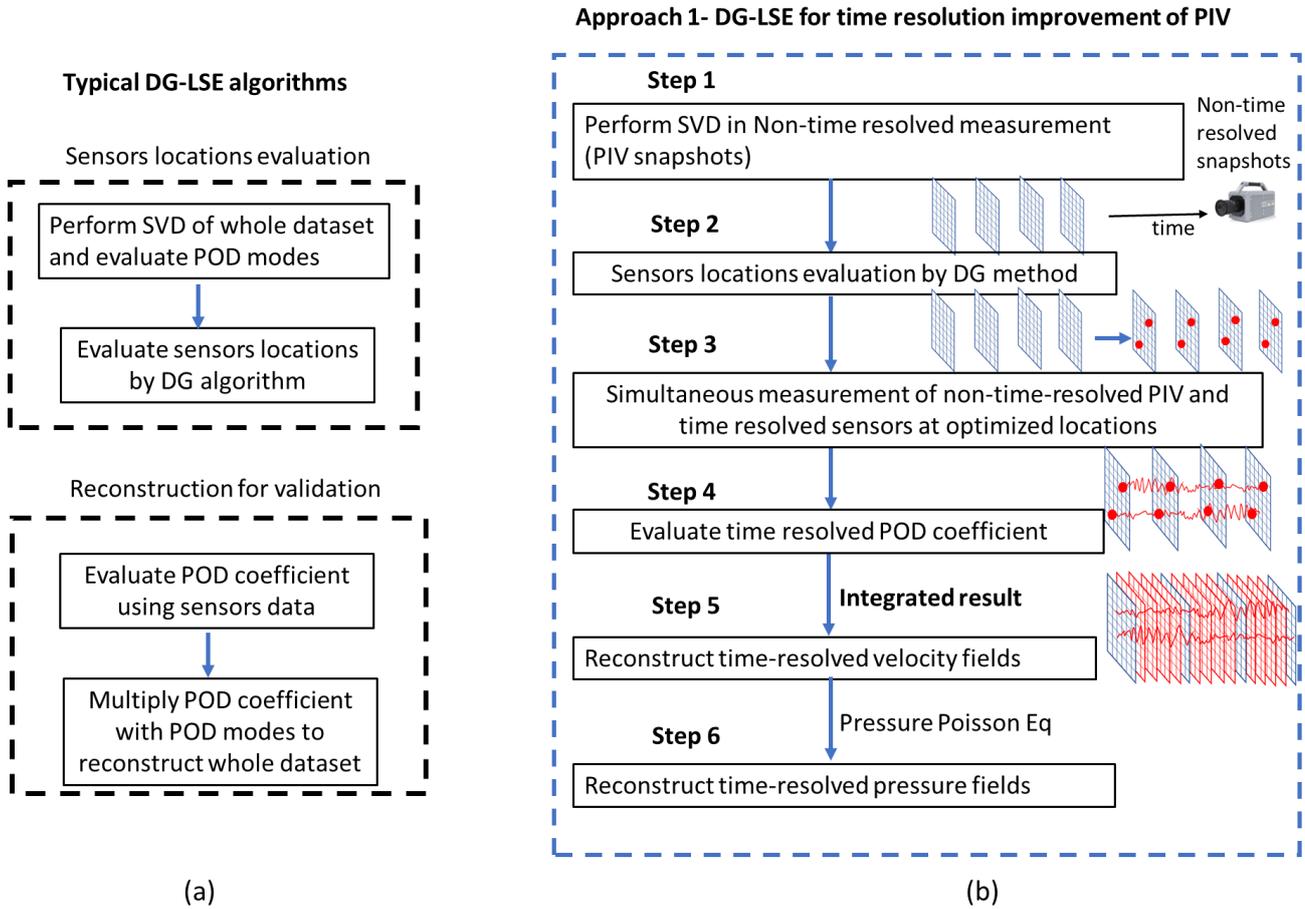

Fig. 3. (a) Typical DG-LSE sensor selction method (b) algorithm of time resolution improvement and pressure field estimation by non-time resolved velocity field and time-resolved point sensors using DG-LSE algorithm

**3.4 Approach 2: direct pressure estimation approach based on sensor selection method (DPE-SSM)**

In this approach, firstly, non time resolved velocity fields are obtained by experiments (Fig. 1). The pressure field is estimated from non-time-resolved velocity data by using pressure Poisson equation/ Navier Stokes equation (step 2). In step 3 the sensors locations are optimized by applying Eq. 10 or Eq. 12 using DG-LSE method depending on the number of sensors. Then physical point sensors can be placed to obtain time resolved data (step 4). Then time resolved POD coefficient can be obtained by Eq.14 (step 5). In step 6 the time resolved pressure fields can be reconstructed by applying from Eq. 15.

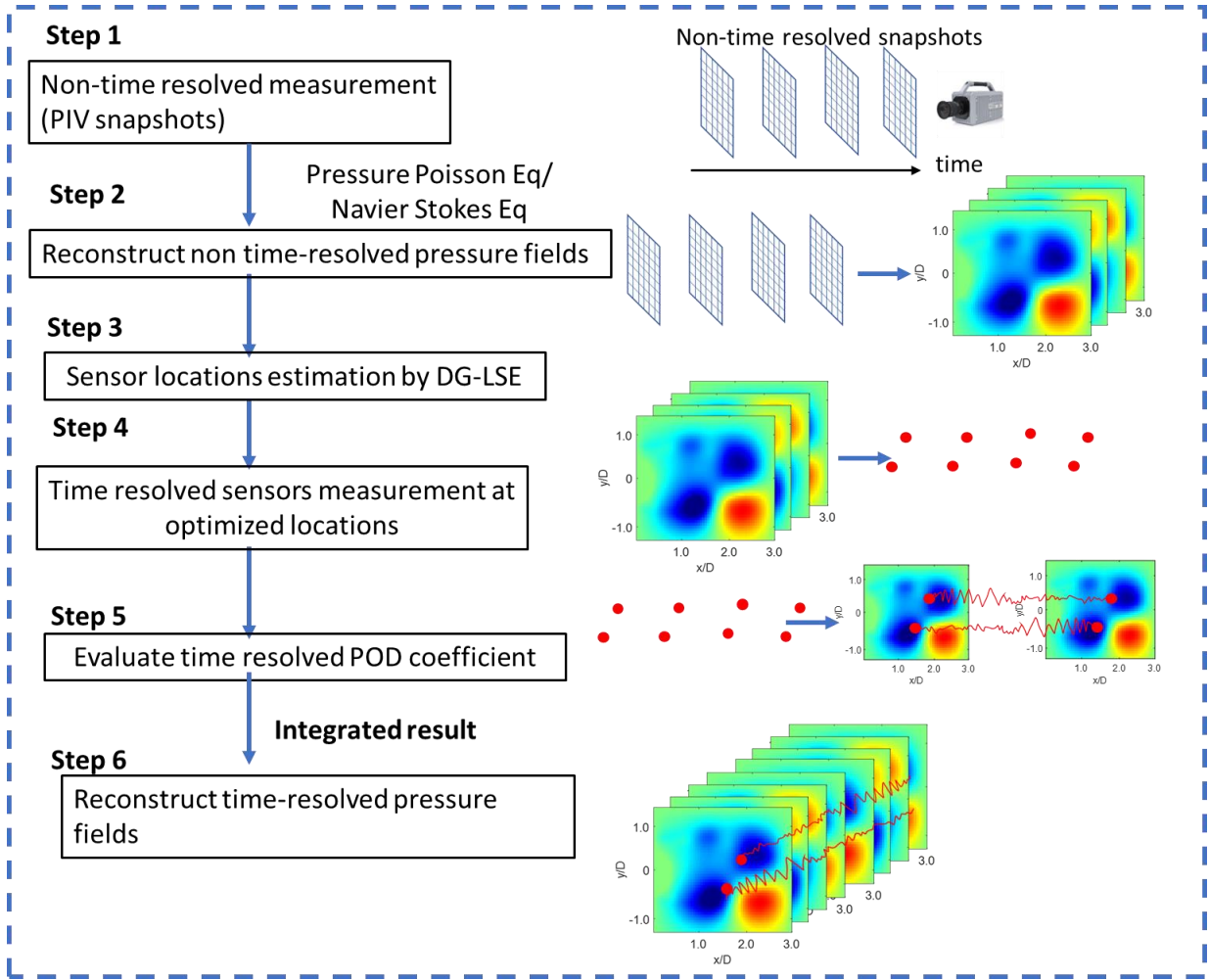

Fig. 4. Steps of approach 2

**3.5 Pressure estimation from velocity data**

In literature, two major methods are reported for the pressure estimation from the measured velocity data: Navier Stokes equations with path integration and the pressure Poisson equation with appropriate boundary conditions. Relation between the pressure and the velocity in the Navier-Stokes equations can be written as

$$\nabla p = -\rho \frac{Du}{Dt} + \mu \nabla^2 u \qquad (20)$$

Where $u$, $p$, and $\mu$ are the fluid flow velocity vector, the pressure, and the local dynamic viscosity, respectively. The right-hand side terms of the equation are provided by the velocity from the PIV measurement for the estimation of the pressure gradient. In literature, either Eulerian[27] or Lagrangian[28] methods are selected. In the Eulerian approach, material acceleration can be estimated as below:

$$\frac{Du}{Dt} = \frac{\partial u}{\partial t} + (u.\nabla)u \tag{21}$$

In present study pressure Poisson equation is used for pressure field reconstruction. This is derived by taking divergence of Eq. (20):

$$\nabla^2 p = -\rho \nabla.(u.\nabla)u. \tag{22}$$

Two kinds of Neumann boundary conditions are applied at all the boundaries for pressure estimation problem; first, pressure gradient from equation (20) is used and second Taylor hypothesis-based boundary conditions [27].

$$\nabla p = -\rho\{-(\bar{u}.\nabla)u' + (u.\nabla)u - v\nabla^2 u\} \tag{23}$$

$\bar{u}$ is the local ensemble average velocity and $u'$ is the fluctuating part of velocity. These quantities are estimated based on the non-time resolved PIV data so it might not be exact but this is the best way to use the accessible information. The calculation of velocity gradients, time resolution and the non-exact boundary conditions are some of the sources of error propagation into the pressure field. A study on error propagation dynamics in PIV based pressure estimation from both measurement domain and boundaries are well discussed (Pan et al. 2016) (Nie et al. 2022). As discussed by [29], the Pressure Poisson Equation (PPE) and Omnidirectional Integration (ODI) methods are mathematically equivalent when consistent boundary conditions are applied. This demonstrates that the choice between PPE and momentum-equation-based integration does not fundamentally change the reconstructed pressure field, provided that the boundary treatment is consistent and accurate.

In our work, the PPE is not the focus of improvement; it simply provides the reference non–time-resolved pressure field used for sensor optimization and data-driven reconstruction. The proposed method itself is independent of whether the reference field originates from PPE or Navier–Stokes-based integration, as it relies on assimilating time-resolved sensor data to recover temporal dynamics. Moreover, as Pryce et al. (2025) discuss, the boundary-condition issue is a common limitation in all pressure-from-velocity formulations, not unique to PPE.

Therefore, our approach is a data-driven framework that can incorporate either PPE- or Navier–Stokes-based reference pressure fields, with the main contribution being its capability to reconstruct time-resolved pressure efficiently using limited sensor data, rather than proposing a new pressure-recovery equation. Both pressure estimation methods have some advantages and disadvantages, but present study does not incline towards any developments of these methods rather it assesses how DG-LSE evaluated time resolved flow fields perform in pressure estimation from pressure Poisson equation.

## 4. Results

### 4.1 Time-resolved velocity field reconstruction

As described earlier, the measured non-time-resolved velocity field from the PIV experiment is decomposed into the POD modes using SVD. Eigenvalues are employed to calculate the energy of the POD modes. Figure 5 shows the cumulative energy associated with the top 100 POD modes. Here, the first five dominant POD modes contain 77.3% of total perturbation energy (equivalent to 97% of the total kinetic energy) of the flow. For the demostration of present methodology for time resolution improvement of dominant structures of PIV data, we have chosen first five modes i.e., $r=5$. For different flow problems, the number of POD modes can be accordingly chosen depending on the desired accuracy.

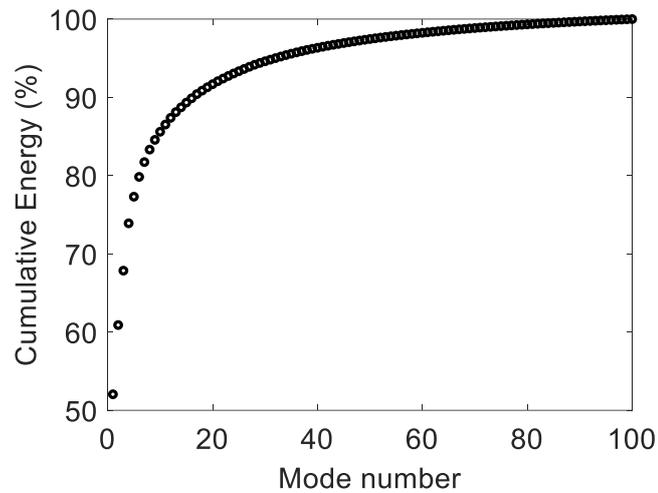

Fig. 5. Cumulative perturbation energy in first 100 POD modes

Figure 6 shows the fluctuating part of the $x$ component of the velocity data ($u$) at three different time instants, reconstructed by considering the first five dominant POD modes of flow. These instants are chosen where non-time-resolved PIV data was not available. These data are referred to as the time-resolved PIV data. The sampling rate of PIV measurement in this case was 50 Hz.

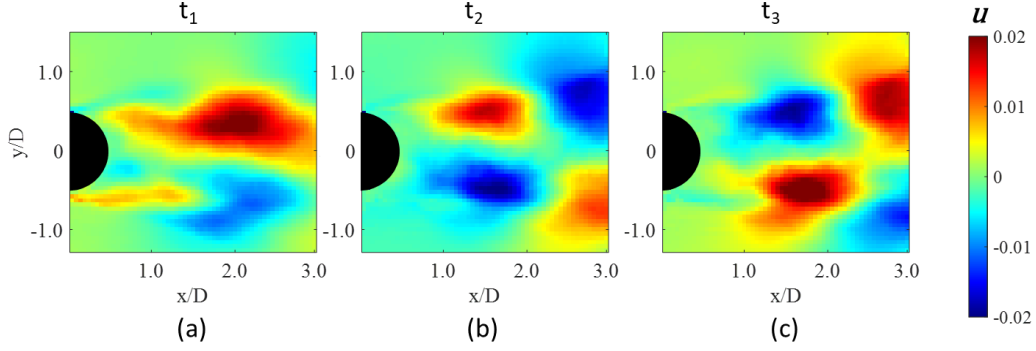

Fig. 6. *x*-component of the time-resolved PIV velocity field in the wake of the cylinder

The number of POD modes and number of sensors are independent parameters to be chosen by the user depending on desired accuracy and cost of sensors. Once number of POD modes are selected, the method presented in section 3 optimizes the location of the chosen number of sensors and time resolved PIV data is reconstructed with different number of sensors. Then reconstructed error is evaluated by comparing reconstructed time resolved PIV data and time-resolved PIV data. Fig. 7 shows the effect of the number of sensors on the reconstruction error with optimized and unoptimized sensors (Eq. 13 to 15 were used for reconstruction), $Error\_tr$ (RMS error over the whole domain and over all 3500 snapshots) which is defined as below:

$$\text{Error\_tr} = \sqrt{\frac{1}{M}\frac{1}{N}\sum_{m=1}^{M}\sum_{n=1}^{N}\left(\left|\frac{\widehat{X}_{tr\_m,n}}{\max(X_{tr})} - \frac{X_{tr\_m,n}}{\max(X_{tr})}\right|^2\right)} \tag{24}$$

Here, $N$ is the total number of snapshots, $n$ is the time instant, $M$ is the total number of data points in a flow field at an instant, $\widehat{X}_{tr}$ is estimated time-resolved truncated velocity data at an instant by sensor selection method, and $X_{tr}$ is time-resolved truncated time-resolved PIV velocity data. The reconstruction error by optimized sensors with DG-LSE methods is reduced with increasing number of sensors and with unoptimized sensors the errors are not much afftected by the number of sensors. This shows that using unoptimized sensors fails to leverage the information available in time resolved sensor data. However, we also want to minimize the number of sensors in the flow domain due to its intrusive nature and reduce the cost of the experiment. So, a decision must be made by considering the cost and intrusiveness versus the accuracy desired.

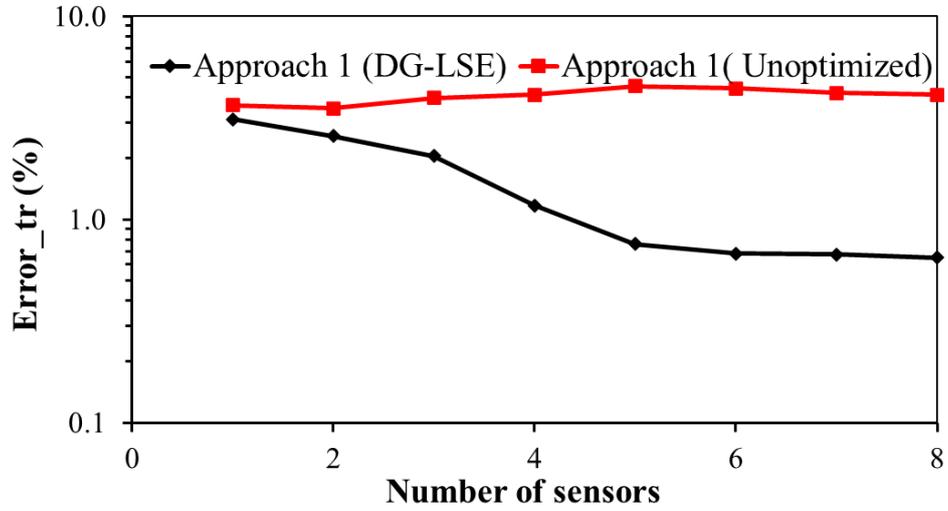

Fig. 7 The reconstruction RMS error of optimized and unoptimized sensors by approach 1 with the increasing number of sensors

Figure 8 presents the velocity field evolution in the time estimated by the present algorithm with four sensors selected by the DG-LSE method. The velocity fields are at the same time instances as in Fig. 6. The locations of optimized sensors are indicated by the open circles in the corresponding velocity field. The sensor locations are enclosed in circles for clearer illustration. In the flow field, out of 3013 candidate sensor points (points where velocity measurements are available), only four points are used for time-resolved sensor measurement. The velocity fields reconstructed using the sensors selected by the DG-LSE method are qualitatively accurate, as seen by comparing Fig. 6 and 8.

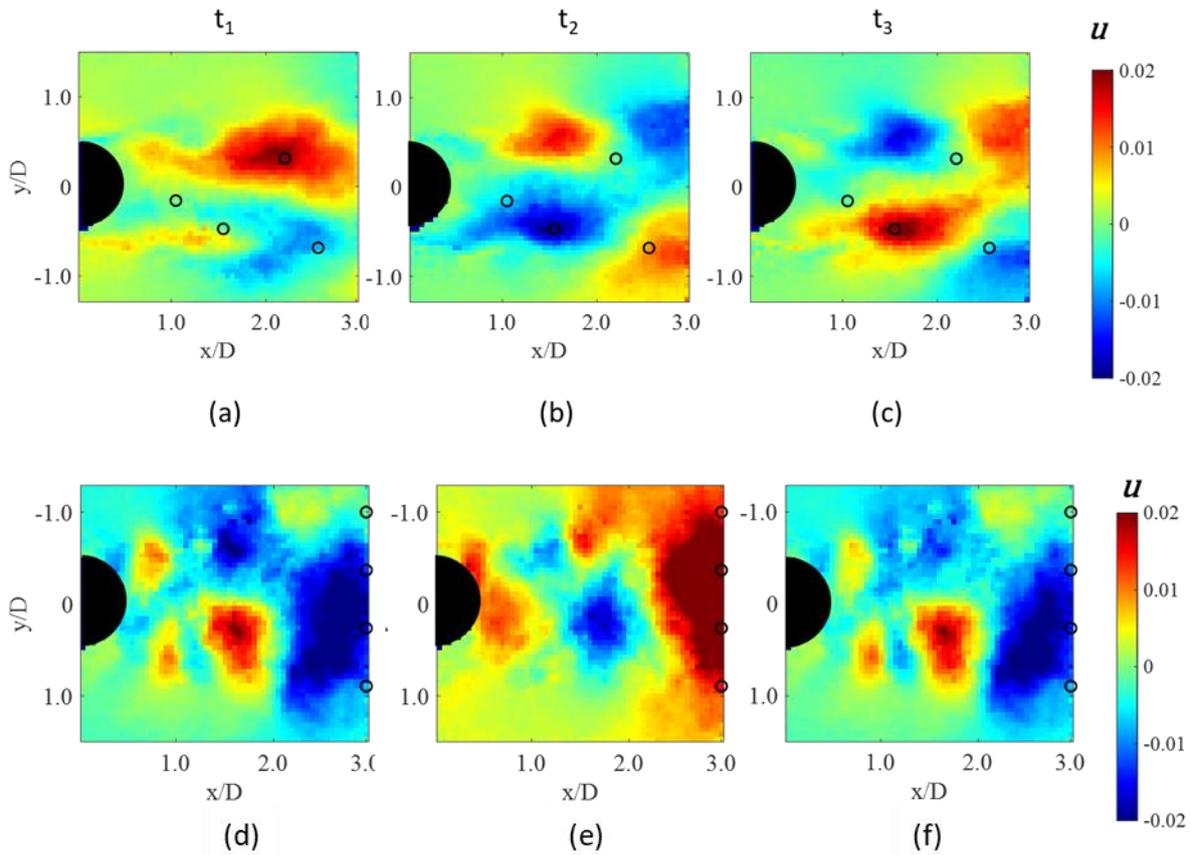

Fig.8 Reconstructed x-component of velocity data by approach one using DG-LSE sensor selection method with optimized sensors (a-c) and unoptimized sensors (d-f)

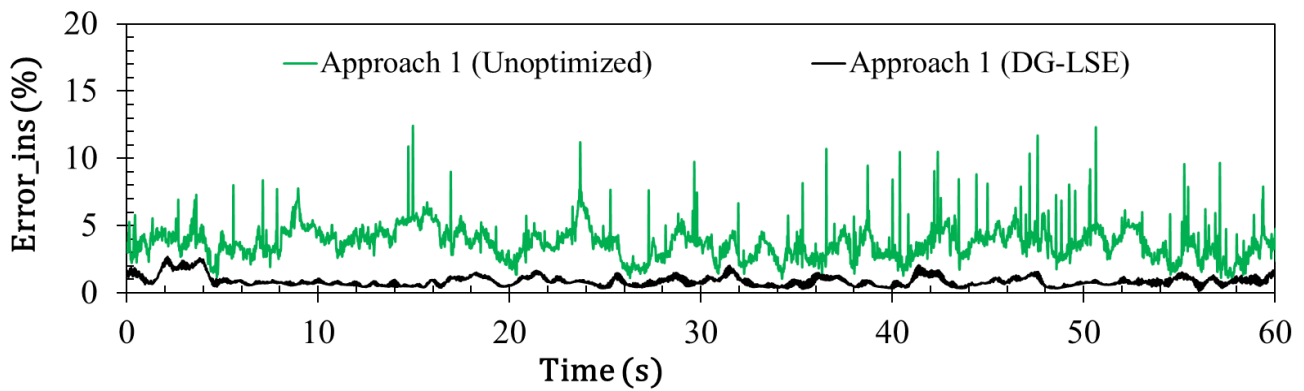

Fig.9 Time series of RMS reconstruction error in velocity estimation by optimized and unoptimized sensors with approach one with DG-LSE method

Further insights into the capability of the DG-LSE method in time resolution improvement of PIV can be gained by looking at the overall time history of error. Figure 9 illustrates the time evolution of error in the instantaneous flow field reconstruction. The root mean square errors of instantaneous snapshots (RMS error in whole domain) using four optimized selected by the DG-LSE (approach 1 ) and unoptimized sensors are shown. The definition of instantaneous reconstruction error is as below:

$$\text{Error\_ins}(n) = \sqrt{\frac{1}{M} \sum_{m=1}^{M} \left( \left| \frac{\widehat{X}\_tr_{m,n}}{max(X\_tr)} - \frac{X\_tr_{m,n}}{max(X\_tr)} \right|^2 \right)} \qquad (25)$$

Approach 1 with optimized sensors shows under 3% reconstruction error in the instantaneous velocity field however in with unoptimized sensors this error this error went upto approximately 15%.

**4.2 Pressure field reconstruction**

Once the time-resolved velocity field is available, the time-resolved pressure field can be estimated using method presented in section 3.5. Pressure Poisson equation is used for pressure estimation from truncated time resolved velocity data (measured time-resolved PIV data) which are shown in Fig. 10 (a-c) and they are referred as the time-resolved PIV-based pressure field because they were estimated from time resolved velocity data of PIV. The pressure is estimated by an iterative method, and iteration is repeated until the root mean square error becomes of the order of $10^{-5}$. Two kinds of Neumann boundary conditions, one, based on Navier Stokes (equation 20) is applied at all the walls in both approach 1 and approach 2.and other based on Taylor hypothesis (equation 23) [27] is applied at all the walls in approach 2.

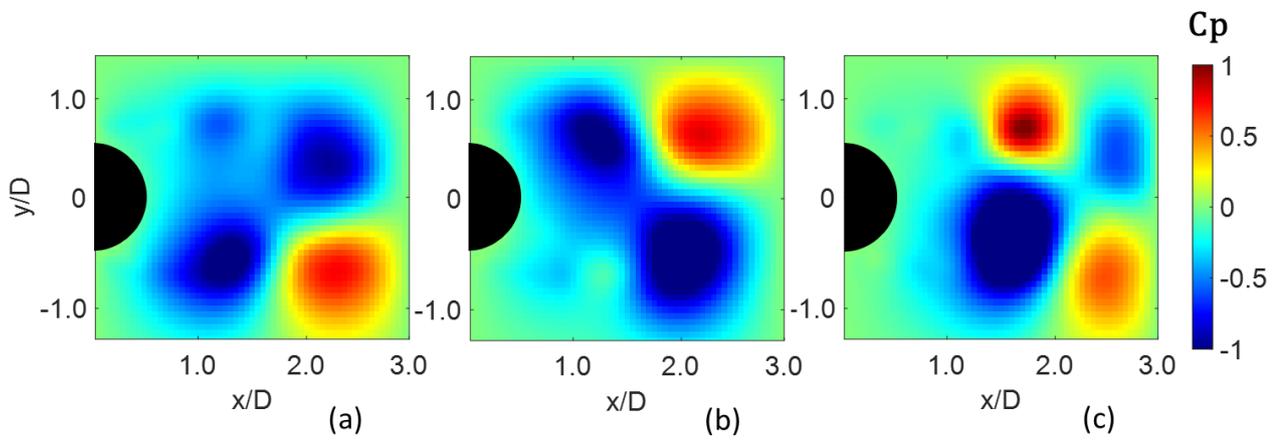

Fig.10 (a-c) pressure field evaluated from pressure Poisson equation corresponding to velocity vectors at three instances

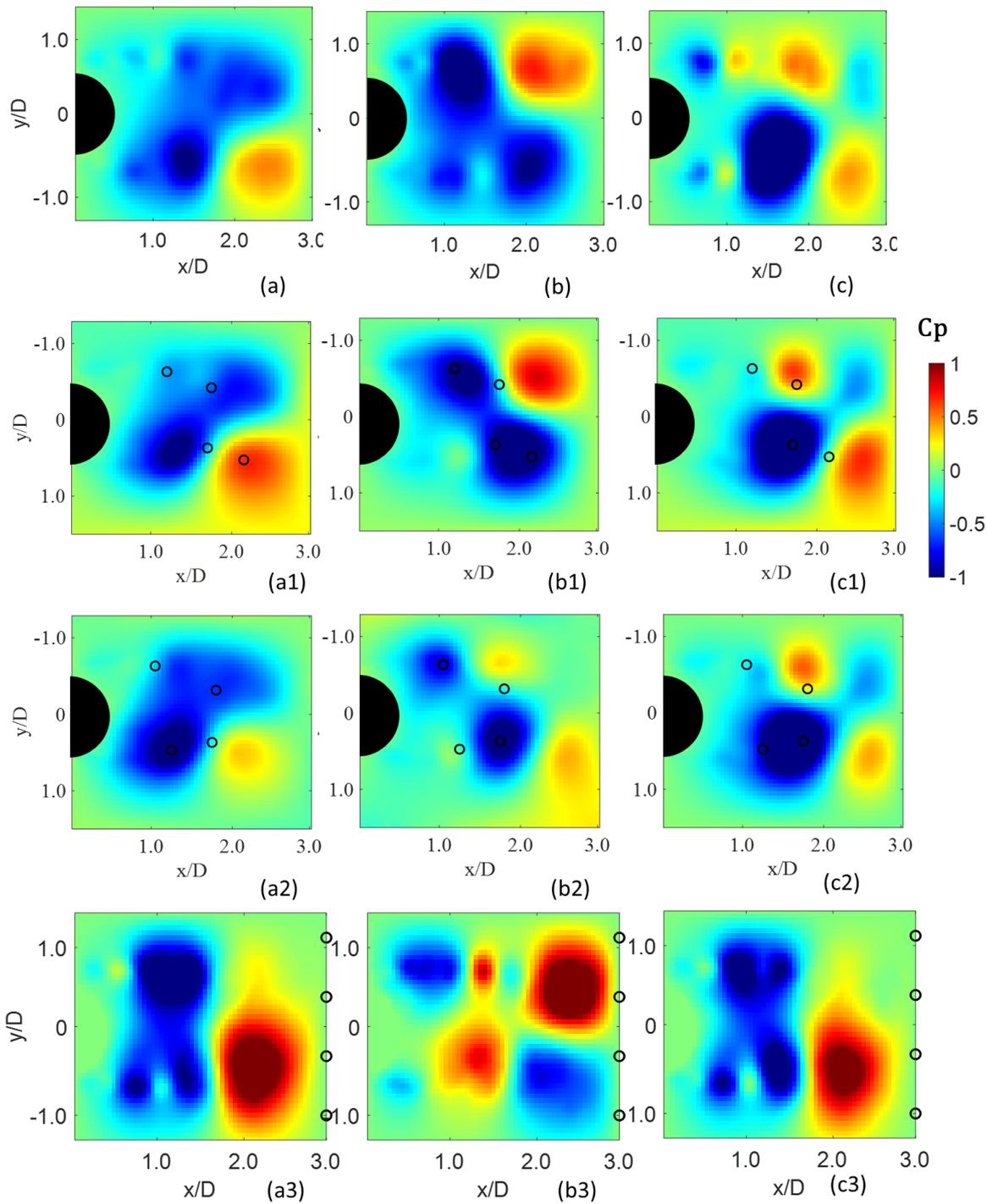

Fig.11 Reconstructed time-resolved pressure field evaluated from pressure Poisson equation by approach 1 (a to c), (a1 to c1) approach 2 with Neumann boundary condition (a2 to c2) approach 2 with Taylor hypothesis (a3 to c3) approach 2 with unoptimized sensors

Figure 11 (a to c), shows the reconstructed time-resolved pressure fields evaluated by DG-LSE method using approach 1 with four sensors at three instances. The reconstructed pressure fields are not the same as time-resolved PIV-based pressure fields (Fig. 10). This indicates that instantaneous pressure field reconstruction is very sensitive to uncertainty in velocity fields. However, low pressure and high-pressure regions in flow are comparable to corresponding time-resolved PIV-based pressure field. The error in the pressure field will be larger than that in the velocity field due to the propagation of errors. These errors come from errors in calculating source terms of the pressure Poisson equation, which originate from the uncertainty of measured velocity data [30,31]. Figure11 (a1 to c1) is the pressure reconstructions at instants using approach 2 with four sensors with Neumann boundary condition. This result show that with four sensors, approach 2 works better than approach 1. Figure 11 (a2–c2) presents the reconstructed pressure fields obtained using Taylor hypothesis based boundary condition and it can be seen that approach 2 with Neumann boundary condition works better than with Taylor hypothesis based boundary conditions. In practice, both kinds of boundary conditions can be tried and one that works best for a particular kind of flow can be utilized. In Fig. 11 (a3-c3) the pressure fields are reconstructed with approach 2 with unoptimized sensor locations. Qualitatively, these results show poor similarity with the reference pressure fields shown in Figure 10(a–c). The effect of unoptimized sensor placement is therefore highly significant, leading to a severe degradation in reconstruction accuracy of the pressure field.

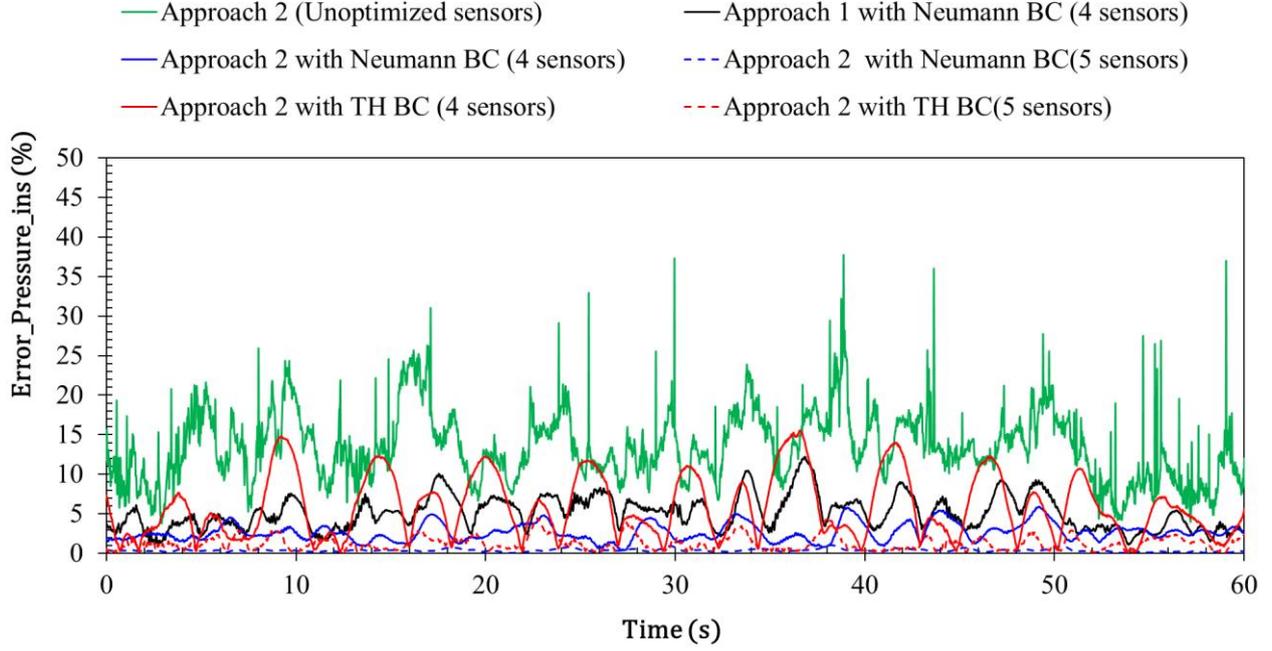

Fig. 12 Time series of reconstruction RMS error in pressure estimation by optimized sensors with DG-LSE method by approach 1 with four sensors (with optimized), approach 2 with four and five unoptimized sensors sensors

Consider the definition of instantaneous (at an instant $n$) RMS error in reconstructed pressure fields in whole domain and over 3500 snapshots as below:

$$\text{Error\_Pressure\_ins}(n) = \sqrt{\frac{1}{M}\sum_{m=1}^{M}\left(\left|\frac{\widehat{X}\_Pressure_{tr\_m,n}}{max(X\_Pressure_{tr})} - \frac{X\_Pressure_{tr\_m,n}}{max(X\_Pressure_{tr})}\right|^2\right)} \quad (26)$$

Here $X\_Pressure_{tr}$ is time resolved pressure field data evaluated from PIV data measured at 50 Hz sampling rate. $\widehat{X}\_Pressure_{tr\_m,n}$ is time resolved pressure data evaluated from estimated time resolved data by DG-LSE method. Figure 12 shows the time series of instantaneous pressure reconstruction errors for Approach 1 (with Neumann boundary condition with four optimized) and Approach 2 (with four and five optimized sensors with Neumann and Taylor hypothesis based boundary conditions and four unoptimized sensors). The results indicate that the estimated pressure error using five sensors in Approach 2 is significantly lower than that obtained with four sensors in either approach. In contrast, when unoptimized sensors are used, the pressure error reaching values of up to 40%.

To study the effect of number of sensors on reconstruction of pressure data in both approach 1 and 2 the rms errors in the pressure field are shown in Fig. 13. Fig. 13 shows the reconstruction error (Eq. 27) for several different numbers of sensors.

$$\text{Error\_Pressure} = \sqrt{\frac{1}{N}\frac{1}{M}\sum_{n=1}^{N}\sum_{m=1}^{M}\left(\left|\frac{\widehat{X}\_Pressure_{tr\_m,n}}{max(X\_Pressure_{tr})} - \frac{X\_Pressure_{tr\_m,n}}{max(X\_Pressure_{tr})}\right|^2\right)} \qquad (27)$$

An increase in the number of sensors leads to the improvement in the estimation of the velocity fields (Fig.7) and, consequently, the pressure fields using approach 1 and 2, as shown in Fig. 13. Approach 1 shows lowest error among all approaches when only one sensor is used and approach 2 with Neumann boundary condition has lower error when number of sensors is increases from 2 to 8. Significant reduction in error is observed when number of sensors increases from four to five in approach 2 with both boundary conditions. At this condition the number of sensors is the same as the number of POD modes. It can be pointed that for more complicated flows, where a greater number of POD modes are required for approximate reconstruction, more sensors in optimal locations will be needed for time resolution improvement. This kind of significant error reduction was also observed in previous work on time resolution improvement of ultrasonic velocity profiler using extended POD with sensor selection method [9]. On the other hand, when unoptimized sensors are used, there is no noticeable improvement in reconstruction accuracy, even with an increased number of sensors, indicating that sensor placement plays a more critical role than sensor quantity in achieving accurate field reconstruction. After applying both boundary-condition treatments, we found that Approach 2 remains robust, because the optimized, time-resolved sensors compensate for the lack of temporal information in the non–time-resolved PIV data. With five optimally placed sensors, the reconstructed pressure field agrees very well with the reference time-resolved pressure for both boundary-condition formulations as shown in Fig. 12. It is important to note that this level of accuracy is achieved only with optimally placed sensors but random sensor placement does not yield comparable performance.

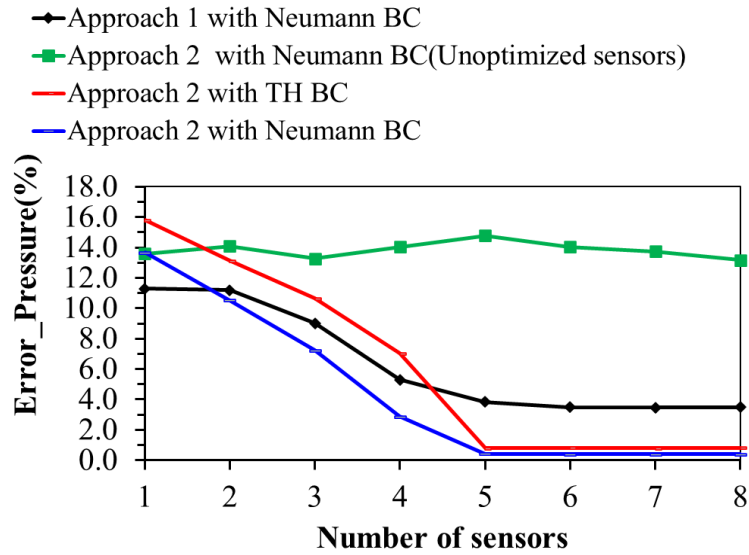

Fig. 13 Root mean square reconstruction error of pressure field estimated by approach 1 with optimized sensors and approach 2 with optimized and unoptimized sensors

Table 1 presents computing time requirements for various steps involved to evaluate time resolved pressure field from non-time resolved PIV data in approach 1(optimized) and 2(optimized and unoptimized). The computing system for these calculations is 11th Gen Intel(R) Core(TM) i7-1195G7 @ 2.92 GHz, 16.0 GB with 64-bit operating system. Total time required for time-resolved pressure fields estimation is 2989.66 Sec in approach 1 with optimized sensors and 98.53 Sec in approach 2. Approach 2 is approximately thirty times faster than approach 1 with optimized sensors. This is expected because the number of pressure Poisson equation calculations are significantly less for second approach. The difference in time requirements between approach 1 and 2 will increase with the increase in time resolution improvement needed. In case, both time-resolved velocity fields and pressure fields are required, approach 2 will require 112.06 Sec which is still 26.67 times less than approach one.

Table 1 Time requirement of various steps of approach 1 with optimized and unoptimized sensors and approach 2

| Approach 1 | | | Approach 2 | | |
|---|---|---|---|---|---|
| | Optimized | Unoptimized | | Optimized | Unoptimized |
| Steps | Time requirement | | Steps | Time requirement | |
| Optimize location of sensors in non-time-resolved velocity data | 1.13 Sec | 0 | Non-time-resolved pressure field estimation from non-time-resolved velocity data | 39.88 Sec | 39.88 Sec |
| Non-time-resolved to time resolved velocity by DG-LSE method | 13.53 Sec | 13.53 Sec | Optimize the location of sensors from non-time resolved pressure fields | 1.13 Sec | 0 Sec |
| Time-resolved pressure fields estimation from non-time-resolved pressure fields | 2975 Sec | NA | Time-resolved pressure field from non-time resolved pressure fields and time resolved sensors | 57.52 Sec | 57.52 Sec |
| Total | 2989.66 Sec | NA | Total | 98.53 Sec | 97.4 Sec |

In this work, we have considered a case where sensor can be placed at every point where PIV data is available. However, it is possible that in actual experiments, it is infeasible to place sensors at these locations. In that case, a constrained sensor selection problem can be solved where the infeasible locations are omitted. After an optimal sensor location is obtained for the constrained problem, the steps from current technique can be implemented. It is expected that this technique will perform better than the case with random sensor placement.

The present study introduces three novel contributions. First, the location evaluation step of a sensor selection framework is used to optimize sensor positions based on a non-time-resolved pressure field, which is computed from non-time-resolved PIV velocity data using the pressure Poisson equation. This approach differs from previous studies, which typically focus on velocity field reconstruction and do not optimize sensor placement specifically for enhancing the temporal resolution of velocity or pressure fields. Second, the reconstruction step of the sensor selection method is applied to estimate time-resolved pressure fields by integrating sparse, time-resolved pressure measurements acquired at the optimized sensor locations with the non-time-resolved pressure field. This enables the reconstruction of time-resolved pressure dynamics without requiring time-resolved velocity input, offering a significant advantage in experimental scenarios where high-speed PIV measurements are impractical or unavailable. Third, since the pressure field is estimated from a reduced number of non-time-resolved PIV snapshots, the computational effort for pressure estimation is substantially lower, making the method efficient when processing large datasets.

## 5. Conclusions

The time resolution of the velocity and pressure fields are improved by non-time-resolved PIV data and sensors data where the locations of sensors are optimized using sensor selction methods. Two approaches 1 and 2 are discussed and compared with regards to the accuracy of final pressure fields and computing time requirements. Approach 1 first reconstructs time resolved velocity field from non-time resolved PIV data and optimized sensor data and then uses pressure Poisson equation to estimate time resolved pressure fields. Approach 2 first estimates non-time resolved pressure field from non time resolved PIV data then reconstructs the time resolved pressure field using optimized time resolved sensor data. The reconstruction error of time resolved pressure field reduced significantly when number of sensors becomes equal to POD modes in approach 2. Approach 2 is thirty times faster than approach 1 in estimation of time resolved pressure field from non time resolved velocity fields when the time resolution is improved fifty folds. Approach 2 solves pressure poisson equation with non-time resolved velocity field whereas approach 1 solves pressure Poisson equation with time resolved velocity fields. Hence, the number of pressure Poisson solutions required in approach 2 are lower which makes it faster.

**Acknowledgements**

This work was financially supported by Indian Institute of technology Hyderabad with grant number SG-193 under seed grant.